\documentclass[12pt]{iopart}

\usepackage{iopams}
\usepackage[english]{babel}
\usepackage{graphicx}
\usepackage{cite}
\usepackage{color}
\usepackage{bm}
\usepackage{multirow,color,relsize,ulem,microtype}

\newcommand{\be}{\begin{equation}}
\newcommand{\ee}{\end{equation}}

\newcommand{\bra}[1]{ \langle #1|}
\newcommand{\ket}[1]{ |#1\rangle}
\newcommand{\prodesc}[2]{ \langle #1|#2\rangle}

\def\ba{\begin{array}}
\def\ea{\end{array}}

\begin{document}

\title{Symmetry in optics and photonics: a group theory approach}

\author{B. M. Rodr\'iguez-Lara}
\address{Tecnologico de Monterrey, Escuela de Ingenier\'ia y Ciencias,  Monterrey 64849, M\'exico,  \\
Instituto Nacional de Astrof\'isica, \'Optica y Electr\'onica,   Puebla CP 72840, M\'exico}
\ead{bmlara@itesm.mx}

\author{Ramy El-Ganainy}
\address{Department of Physics and Henes Center for Quantum Phenomena, Michigan Technological University, Houghton MI 49931, USA}

\author{Julio Guerrero}
\address{Departamento de Matem\'aticas, Facultad de Ciencias Experimentales y de la Salud, Campus Las Lagunillas, Universidad de Ja\'en, 23071 Ja\'en, Spain.\\
Departamento de Ingenier\'{\i}a y Tecnologr\'{\i}a de Computadores, Facultad de Inform\'atica, Campus Espinardo, Univesidad de Murcia, 30100 Murcia, Spain.}

\begin{abstract}
	Group theory (GT) provides a rigorous framework for studying symmetries in various disciplines in physics ranging from quantum field theories and the standard model to fluid mechanics and chaos theory. To date, the application of such a powerful tool in optical physics remains limited. Over the past few years however, several quantum-inspired symmetry principles (such as parity-time invariance and supersymmetry) have been introduced in optics and photonics for the first time. Despite the intense activities in these new research directions, only few works utilized the power of group theory. Motivated by this status quo, here we present a brief overview of the application of GT in optics, deliberately choosing examples that illustrate the power of this tool in both continuous and discrete setups. We hope that this review will stimulate further research that exploits the full potential of GT for investigating various symmetry paradigms in optics, eventually leading to new photonic devices.
\end{abstract}
\maketitle

\section{Introduction}
Symmetry principles play a crucial role in modern physics. Historically, one can trace interest in symmetry concepts to the early works on platonic solids by the ancient Greeks. In modern time, one of the first investigations that brought the notion of symmetry to the forefront of physical science was the discovery by Emmy Noether \cite{Noether1918p235} that conservation laws and continuous symmetries are connected; e.g., the conservation of energy, linear and angular momenta are a direct outcome of temporal, spatial and rotational symmetries, respectively. Almost at the same time, the concept of gauge invariance was introduced by Herman Weyl in an attempt to unify gravity with electromagnetism. In quantum physics, the concepts of invariance under particle permutation (without or with a change of sign) led to the discovery of elementary particles classifications into bosons and fermions together with the associated particle statistics (Bose-Einstein or Fermi-Dirac). Despite these efforts, it was not until the seminal work of Wigner that group theory (the mathematical tool for studying symmetry) was formally employed in physics – particularly to study how symmetries of molecular configurations affect their vibrational spectra. Nowadays, group theory is an integral part of physics with applications ranging from quantum field theories, atomic and nuclear physics to quantum information and chaos theory to just mention few examples.

In optics, the tremendous progress over the past few decades have benefited a little from group theoretical techniques, except for a few notable examples (see Ref. \cite{Wolfbook} and references therein as well as  Refs. \cite{Khan,Vance,Vance2, PBG, RJLR}). Recently however, new research directions that exploits how quantum-inspired symmetries can be used to engineer novel optical structures have emerged. These include for instance PT symmetry (and non-Hermiticity in general) \cite{Bender1,CPA,CPALaser,Constant_Intensity,EP_PRL,EP_exp,El-Ganainy_OL07,Feng2,Guo,Hodaei,Kostas_Anamolous,Kottos_RLC,Lin,Musslimani_prl08,RE_EPs,RE_Nonlinear,RE_Optomechanic,Rotter_EP,phonon_laser,Yang,Makris_prl08,Longhi,Loss,Ruter, Walk,Schomerus1,Schomerus2,Wiersig,Teimourpour, Xiao, Schomerus_PTLaser} and supersymmetry \cite{Cooper,Wolf,BML_SUSY, SUSY_scattering, Miri_SUSY,Longhi_SUSY,RE_SUSY1,RE_SUSY2,RE_SUSY3,RE_SUSY4,Park_SUSY,RE_SUSY5}. Despite recent intense activities in these fields, group theory remains largely outside the scope of these studies.
In this article, we try to bridge this gap (at least partially) by presenting a brief review over some of the applications of group theory in continuous and discrete photonic systems in order to demonstrate its power with the hope that this will encourage further research in this largely unexplored area of research. For sake of clarity, we present more details on some of the technical mathematical terms used here in the supplementary material.

In general, systems admitting analytic solutions are rare, and when they exist, they have an underlying
symmetry that can be exploited by group theory to obtain their solutions. The most important method for this purpose is Lie's method to compute symmetries of differential equations (see Section 2).
But a most fruitful approach is to design differential equations with prescribed (well-known) symmetries  and use
group-theoretical methods for obtaining their solutions (see  Section 3).

Particularly, we focus on two specific examples: (1) The symmetry group associated with Helmholtz equation and how it can be used to derive new non-diffracting solutions in free space; and (2) The application of group theory in discrete photonic systems.

\section{Euclidean symmetry group of the Helmholtz equation}
In this section, we review how group theory can be applied to obtain special solutions of the scalar Helmholtz equation; in particular, we will focus on propagation invariant beams.
In general, light propagation is described by Maxwell equations together with their constitutive relations.
For monochromatic fields in homogeneous, isotropic, and linear media, these take the form of the vector Helmholtz equation:
\begin{equation} \label{eq:Eq1}
\left(\bm{\nabla}^2 + \beta^2 \right) \bm{E}  = 0.
\end{equation}
The squared wavenumber $\beta^2 = \vec{\beta} \cdot \vec{\beta} = \left( 2 \pi / \lambda \right)^{2}$ is given in terms of the wavelength $\lambda$; and the symbol $\bm{\nabla^2}$ stands for the vector Laplace operator.
For well-defined, homogeneous, and isotropic polarized vector fields, this vector field equation reduces to the scalar Helmholtz equation describing scalar waves.

In addition, it is well-known that Hertz potential formalism  \cite{Stratton1941} allows the following construction of vector field solutions to Maxwell equations:
\begin{equation} \label{eq:Eq2}
\bm{E} = c_{\rm TE} \left( \nabla \times \hat{u} \Psi \right) + c_{\rm TM} \left( \frac{1}{k} \nabla \times \nabla \times \hat{u}  \Psi \right),
\end{equation}
from a scalar wavefunction $\Psi(\bm{r})$ that solves the scalar Helmholtz equation:
\begin{eqnarray} \label{eq:Eq3}
\left( \nabla^2 + \beta^2 \right) \Psi = 0 ,
\end{eqnarray}
where the unit vector $\hat{u}$ is usually taken along the $z$-axis for propagation invariant fields \cite{VolkeSepulveda2006p867}, the complex coefficients $c_{\rm TE}$ and $c_{\rm TM}$ define the transverse electric and transverse magnetic components of the optical field while the symbol $\nabla^2$ stands for the scalar Laplace operator.

The above discussion demonstrates the central role of the scalar wave equation in electromagnetics and optics:
it is possible to construct any given electromagnetic vector field using a scalar wave solving the scalar Helmholtz
equation and a particular unit vector via Hertz potential formalism. This in turn necessitates a deep understanding of its mathematical structure. This can be best accomplished by employing the group theoretical program developed by Sophus Lie \cite{Lie1880p441} for studying symmetries of differential equations.
Before we proceed, we note that under certain conditions, Helmholtz equation can be further reduced to the paraxial equation of diffraction, which is isomorphic to Schr\"odinger equation, and often used to describe optical beams. While in this review we focus on the symmetry groups of Helmholtz equation, the symmetries of continuous Schr\"odinger equation are also very well characterized and we refer the interested reader to Chapter 2 in Ref. \cite{Miller1984} for details. Also, in the last Section we discuss discrete photonic systems obeying equations which are formally identical to the discrete Schr\"odinger equation.

To this end, we note that first-order symmetries of differential equations (having either ordinary or partial derivatives) are first-order differential operators $\hat{L}$ mapping solutions of the differential equations into other solutions; i.e., if $\Psi$ is a solution of the differential equation, then $\hat{L}\Psi$ is
also a solution.
If $L$ and $L'$ are first-order symmetries, an arbitrary linear combination $\alpha L + \beta L'$  is also a
first-order symmerty (therefore they constitute a vector space), the product (composition of differential operators) $LL'$ is a second-order
differential operator that maps solutions into solutions, and therefore is a second-order symmetry. Higher-order symmetries can be built in the same manner.
The particular combination $[\hat{L},\hat{L}']\equiv \hat{L}\hat{L}'-\hat{L}'\hat{L}$ of two first-order symmetries, known as the
commutator, is also a symmetry.
Thus, first-order symmetries naturally arrange themselves into Lie algebras with the commutator as the Lie bracket. We shall denote by
symmetry Lie algebra (or simply symmetry algebra) of a differential equation  to the largest Lie algebra of first-order symmetries.

Lie's program leads to the six-dimensional Euclidean Lie algebra $\mathcal{E}(3)$ as the underlying symmetry algebra of the scalar Helmholtz equation, Eq.(\ref{eq:Eq3}).
A basis for this symmetry algebra is provided by six operators proportional to the components of linear and angular momenta:
\begin{eqnarray}
\hat{P}_{k} = \frac{\partial}{\partial x_{k}} \equiv \partial_{k}, \quad
\hat{J}_{k} = \left( \bm{r} \times \hat{\bm{P}} \right)_{k},
\end{eqnarray}
in that order, with $k=1,2,3$. In the above we have used the three-dimensional position vector $\bm{r} = (x_{1},x_{2},x_{3})$ and the total linear momentum $\hat{\bm{P} }= (\hat{P}_{1},\hat{P}_{2},\hat{P}_{3})$.
It is straightforward to show these operators fulfill the standard Lie bracket commutation relations:
\begin{eqnarray}
\left[ \hat{P}_{j}, \hat{P}_{k} \right] = 0, \quad \left[ \hat{J}_{j}, \hat{J}_{k} \right] = \epsilon_{jkl} \hat{J}_{l}, \quad \left[ \hat{J}_{j}, \hat{P}_{k} \right] = \epsilon_{jkl} \hat{P}_{l},
\end{eqnarray}
where $\epsilon_{jkl}$ is the Levi-Civita symbol, and the convention of summation over repeated indices has been used. Note that these operators
are generators of the three translations and rotations in $\mathbb{R}^3$.

The use of representation theory of the Euclidean symmetry group $E(3)$, and Miller's program for separation of variables \cite{Miller1975}, which looks for commuting sets of second-order differential operators, quadratic in the Lie algebra operators, naturally allows for deriving the eleven separable solutions for the scalar wave equation \cite{Boyer1976p35,Miller1984}.

To see this, note that the scalar Helmholtz equation is an eigenvalue differential equation in three variables corresponding to the quadratic operator $\hat{P}^2 \equiv \hat{\bm{P}} \cdot \hat{\bm{P}} = \sum_{j=1}^{3} \hat{P}_{j}^2 \equiv \nabla^2$, the Laplacian which is a Casimir of the Lie algebra $\mathcal{E}(3)$, with eigenvalue $- \beta^2$.
According to Miller's program \cite{Boyer1976p35,Miller1984},  there exists two more commuting second-order differential operators, quadratic in the Lie algebra $\mathcal{E}(3)$, in addition to the Laplacian for each coordinate system where Helmholtz equation is separable.
We will call either the operators or eigenvalues invariants  or constants of motion for the scalar Helmholtz equation in the corresponding coordinate system.
The separable scalar wavefunction will also fulfill an eigenvalue equation for these two operators, since the three constants of motion mutually commute.
For example, in Cartesian coordinates, Helmholtz equation can be written in the following form, as already anticipated:
\begin{eqnarray}
\hat{P}^2 \Psi = - \beta^2  \Psi,
\end{eqnarray}
it is straightforward to realize the extra pair of commuting quadratic operators in this representation,
\begin{eqnarray}
\hat{P}_{2}^2 \Psi = - \beta_{2}^2  \Psi, \quad \hat{P}_{3}^2 \Psi = - \beta_{3}^2  \Psi,
\end{eqnarray}
as the Laplacian operator commutes with its components, $\left[ \hat{P}^{2}, \hat{P}_{j}^2 \right] = 0$ with $j=1,2,3$.
The separation constants are $\beta_{2}$ and $\beta_{3}$, the linear  momenta in the directions $x_{2}$ and $x_{3}$, such that the separable solutions in the Cartesian coordinate system,  $x_{1}=x$, $x_{2}=y$, and $x_{3}=z$ is given by,
\begin{eqnarray}\label{cartesian}
\Psi(x,y,z) = {\rm e}^{-{\rm i} \beta_{x} x} {\rm e}^{-{\rm i} \beta_{y} y} {\rm e}^{-{\rm i} \beta_{z} z},
\end{eqnarray}
with $\beta^2 = \beta_{x}^2 + \beta_{y}^2 + \beta_{z}^2$, are plane waves conserving linear momentum.
Note that this is a propagation invariant wave, i.e. it is of the form  $\Psi(\bm{r}) = \Psi_{\perp}(x,y) {\rm e}^{{\rm i} \beta_{z} z}$ and we can use it to construct different optical beams.

It should be stressed that other pairs of commuting quadratic operators like $\{\hat{P}_{1}^2,\hat{P}_{2}^2\}$ or $\{\hat{P}_{1}^2,\hat{P}_{3}^2\}$ could have been chosen. However, these choices lead to the same separation of variables in Cartesian coordinates.
The reason behind this is that different pairs $\{\hat{O}_1,\hat{O}_2\},\{\hat{O}_1',\hat{O}_2'\}$ of commuting quadratic operators related by a similarity transformations by elements $g$ of the Eucliean group $E(3)$, i.e.  $O_i'=gO_i g^{-1}$ for $i=1,2$, leads to the same separation of variables since
similarity transformation preserve the eigenvalues of the operators. Thus, the inequivalent separation of variables are classified by the different orbits under similarity transformation by elements of the Eucliean group $E(3)$ on the set of pairs of commuting  quadratic operators, that turns to be eleven in total \cite{Boyer1976p35,Miller1984}. Hereby, we shall only consider a representative pair in each orbit.

We can construct another three separable solutions that are propagation invariant considering the circular-, elliptic-, and parabolic-cylindrical coordinate systems using the Euclidean symmetry group representation for each of them.
Separation in circular-cylindrical coordinates is provided by the following pair of conserved operators:
\begin{eqnarray}
\hat{P}_{3}^2 \Psi = - \beta_{3}^2  \Psi, \quad \hat{J}_{3}^2 \Psi = - m  \Psi.
\end{eqnarray}
We can take the first as the linear momenta in the propagation direction, and the latter as the component of the total angular momentum in the propagation direction.
It is straightforward to realize that the solution for these eigenvalue equations are Bessel waves:
\begin{eqnarray}
\Psi(r, \theta, z) = J_{m}(\beta_{\perp}^2 r) {\rm e}^{{\rm i} m \theta} {\rm e}^{-{\rm i} \beta_{z} z}  ,
\end{eqnarray}
that have separation variables $\beta_{z}$, a real number, and $m$, an integer, corresponding to the $z$-component of both the lineal and angular momenta, in that order. We have also defined the perpendicular linear momentum $\beta_{\perp} = \beta_{x}^{2} + \beta_{y}^{2}$.
We can use these scalar waves to construct paraxial optical Bessel beams \cite{Durnin1987p1499}, vectorial Bessel beams and even quantize them \cite{Jauregui2005p033411}.

On the other hand, separation in elliptic-cylindrical coordinates is provided by the two conserved operators:
\begin{eqnarray}
\hat{P}_{3}^2 \Psi = - \beta_{3}^2  \Psi, \quad \left( \hat{J}_{3}^2 + f^{2} \hat{P}^{2}_{1} \right) \Psi = a \Psi,
\end{eqnarray}
corresponding, again, to the {$z$}-component of the linear momentum and a composite of the squared $z$-component of the angular momentum and the squared $x$-component of the linear momentum scaled by the squared inter-focal distance of the coordinate system, $f$.
We can think about the latter second order operator as an elliptic momentum.
The separable solutions are given in terms of Mathieu functions,
\begin{eqnarray}
\Psi(\eta, \xi, z) = {\rm e}^{-{\rm i} \beta_{z} z} \left\{ \begin{array}{ll}
\mathrm{Ce}_{n}(\eta, q) \mathrm{ce}_{n}(\xi, q), & n=0,1,2, \ldots \\
\mathrm{Se}_{n}(\eta, q) \mathrm{se}_{n}(\xi, q), & n=1,2, \ldots \\
\end{array} \right.
\end{eqnarray}
with the auxiliary parameter $q = f^2 k_{\perp}^2 / 4$.
Note that we have split Mathieu waves in even and odd as they do not share common eigenvalues for all values of $q$ \cite{Whittaker1927}.
These eigenvalues, or separation constants, are fixed real number for a given value of $n$ and $q$, thus, we can write them as $a \equiv a_{n}^{\rm e}(q)$, for even modes, and  $a \equiv a_{n}^{\rm o}(q)$, for odd modes.

\begin{figure} [t!]
	\center \includegraphics[width=0.75\textwidth]{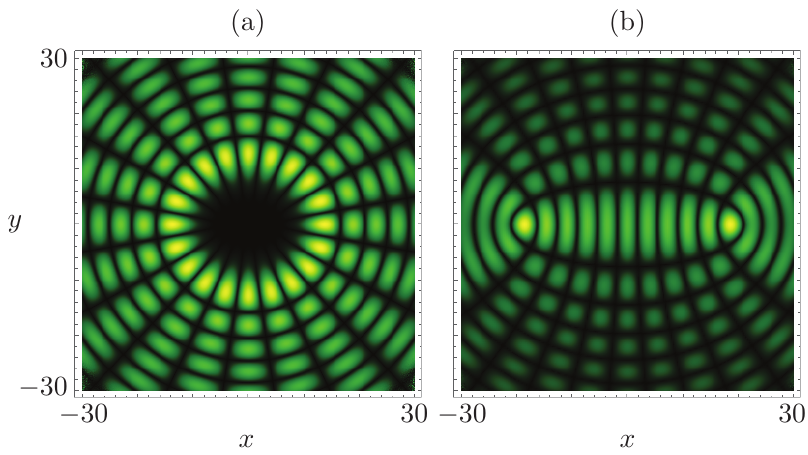}
	\caption{(Color online) Mathieu wave irradiance of even order showing (a) elliptic pattern for parameters $f=5.05$, $q=5$, $n=10$, $a=100.13$, and $b=90.13$; (b) hyperbolic patterns for parameters $f=19.54$, $q=75$, $n=10$, $a=139.44$, and $b=-10.55$.}
	\label{fig:FigMathieu}
\end{figure}
Before moving forward, we note that we can rewrite the elliptic momentum,
$\hat{J}_{3}^2 + f^{2} \hat{P}^{2}_{1} = \hat{L}_{f} - \frac{1}{2} f^2 \left( \hat{P}_{1}^2 + \hat{P}_{2}^{2} \right)$, as a symmetric composition of the $z$-component of the orbital momentum with respect to the elliptic foci in Cartesian coordinates, $\hat{L}_{f} = \left[ (x + f,y,z)  \times \hat{\bm{P}}  \right]_{z} \left[ (x - f,y,z)  \times \hat{\bm{P}}  \right]_{z} + \left[ (x - f,y,z)  \times \hat{\bm{P}}  \right]_{z} \left[ (x + f,y,z)  \times \hat{\bm{P}}  \right]_{z}$,
and the squared transversal lineal momentum.
Interestingly, a positive symmetric composition of the $z$-component of the orbital momentum with respect to the elliptic foci, $\hat{L}_{f} \Psi = b \Psi$ and $b >0$, produces dominant elliptic patterns in the irradiance, Fig. \ref{fig:FigMathieu}a, while negative values, $b < 0$, yields hyperbolic irradiance patterns, Fig. \ref{fig:FigMathieu}b, for both scalar and vector optical Mathieu beams \cite{GutierrezVega2000p1493, RodriguezLara2008p033813}.

\begin{figure}
\center	\includegraphics[width=0.75\textwidth]{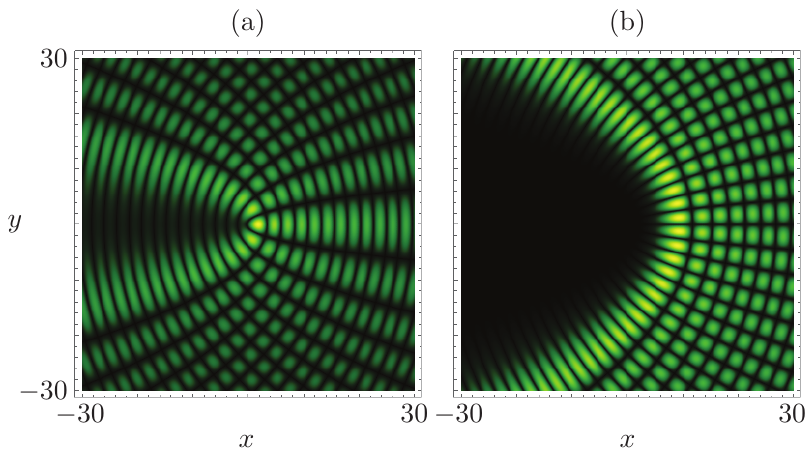}
	\caption{(Color online) Weber wave irradiance of (a) even order for parabolic momentum $a=1.25$ and (b) odd order for parabolic momentum $a=10.5$.}
	\label{fig:FigWeber}
\end{figure}

Finally, separation in parabolic-cylindrical coordinates is provided by the following pair of commuting quadratic operators,
\begin{eqnarray}
\hat{P}_{3}^2 \Psi = - \beta_{3}^2  \Psi, \quad \left( \hat{J}_{3}\hat{P}_{2} + \hat{P}_{2}\hat{J}_{3} \right)  \Psi = a \Psi,
\end{eqnarray}
that are the $z$-component of the linear momentum and the symmetric composition of the $z$-component of the angular momentum with the $y$-component of the linear momentum, that we can call parabolic momentum with real eigenvalue $a$.
These eigenvalue equations yield the even and odd scalar Weber waves,
	\begin{eqnarray}
	\Psi(u,v,z) &=& {\rm e}^{-{\rm i} \beta_{z} z} {\rm e}^{- \frac{\rm i}{2} \beta_{\perp}^{2} \left( u^2 + v^2\right)} \times \nonumber \\
	&& \times \left\{ \begin{array}{ll}
	~_{1}F_{1}\left( \frac{1}{4} -i\frac{b}{2}, \frac{1}{2}, i \beta_{\perp} u^{2} \right) ~_{1}F_{1}\left( \frac{1}{4} -i\frac{b}{2}, \frac{1}{2}, i \beta_{\perp} v^{2} \right),  \\
	u v  ~_{1}F_{1}\left( \frac{3}{4} -i\frac{b}{2}, \frac{3}{2}, i \beta_{\perp} u^{2} \right) ~_{1}F_{1}\left( \frac{1}{4} -i\frac{b}{2}, \frac{3}{2}, i \beta_{\perp} v^{2} \right),
	\end{array} \right.
	\end{eqnarray}
where Kumer confluent hypergeometric function, $_{1}F_{1}(a,b,z)$, has been used \cite{Lebedev1965}.
The conserved parabolic momentum $a$ controls the aperture of the parabolas seen in the irradiance for both even, Fig. \ref{fig:FigWeber}a, and odd, Fig. \ref{fig:FigWeber}, optical Weber waves \cite{RodriguezLara2009p055806}.
These Weber waves can be used to construct scalar beams \cite{Bandres2004p44} and vectorial solutions both in the classical and quantum regimes \cite{RodriguezLara2009p055806}.
Note that the transfer of optical parabolic momentum to cold atom clouds has been studied \cite{RodriguezLara2009p011813R,PerezPascual2011p035303} and shown experimentally for Weber beams \cite{HernandezCedillo2013p023404}.

In addition to propagation invariant fields, where the transversal structure does not change with the propagation length, one can also construct optical accelerating beams (where the transversal structure perpendicular to a curved propagation length is invariant) using the symmetries of the scalar Helmholtz equation with just a change of perspective. If we switch the propagation direction from the $z$-direction to the $y$-direction, then we recover the so-called half-Bessel \cite{Kaminer2012p163901,Mathis2013p2218},  half-Mathieu \cite{Zhang2012p193901} and Weber \cite{Bandres2013p013054} accelerating optical beams from the solutions presented above in circular-, elliptic- and parabolic-cylindrical coordinates, in that order.
Furthermore, spherically symmetric solutions to the scalar Helmholtz equation in parabolic, oblate and prolate spheroidal, and spherical coordinates have been used to construct nonparaxial accelerating waves  \cite{Alonso2012p5175,Bandres2013p30}.


\section{Symmetries in Discrete photonic systems}

In the past two decades, discrete photonics has emerged as a new paradigm for engineering optical structures that exhibit unique properties \cite{Discrete_review}. These systems (often constructed by using waveguide arrays as shown schematically in Fig. 3)  serve as testbed for observing some intriguing phenomena that were first predicted theoretically in the context of condensed matter such as Bloch oscillations \cite{Bessel,Bloch1,Bloch2,Bloch3}, dynamic localization \cite{DL1,DL2}, Anderson localization \cite{AL}  and more recently topological insulators \cite{Topo1,Topo2,Topo3, Topo4}. Additionally, the mathematical analogy between discrete arrays and quantum optics has been also recently investigated \cite{Hector-Christodoulides-Blas,BlasErmakov}.

In this section, we will review some of the applications of group theory in discrete photonic systems described by Coupled Mode Theory (CMT).
But, instead of starting from a differential equation and determine its symmetry
algebra like in previous section when we considered Helmholtz equation, we shall design  differential equations with prescribed (well-known) symmetry algebras and use
group-theoretical methods for obtaining their solutions. For this purpose we shall use as building blocks the number operator (producing a linear propagation constant ramp)
and the step-up and step-down operators (or Susskind-Glogower operators \cite{SG}, producing the coupling between adjacent waveguides) which can be appropriately modulated by functions depending both
in the propagation distance and in the number operator in order to produce the desired symmetry algebra.

%
Particularly, we will discuss the simplest examples of three-dimensional symmetry algebras, namely the Euclidean algebra ${\cal E}(2)$ of translations and rotation in the plane and the semisimple algebras ${\cal SU}(2) \approx {\cal SO}(3)$ and ${\cal SU}(1,1) \approx {\cal SO}(2,1)$.
\begin{figure}
	\centering
	\includegraphics[width=0.65\textwidth] {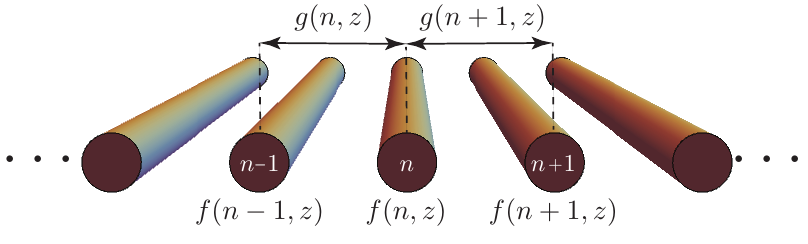}
	\caption{(Color online) Array of coupled waveguides}
\end{figure}

We start by considering a photonic array made of coupled waveguide elements which can be described by the coupled mode formalism:
\begin{equation}
-{\rm i}\frac{{\rm d}{\cal E}_n}{{\rm d}z} =  f(n,z){\cal E}_n+ \left[g(n,z){\cal
	E}_{n-1}+g(n+1,z){\cal E}_{n+1}\right],
\label{Eq}
\end{equation}
with  $n\in\mathbb{Z}$.
If $g(1,z)=0$, this set of equations uncouples into two semi-infinite
sets: $n\leq 0$ and $n\geq 1$.
If further $g(N+1,z)=0$, for $N>0$, it uncouples into three sets: $n\leq 0$,
$1\leq n \leq N$ and $n>N$.

In the general case, the equations can be written as:
\begin{equation}
-{\rm i}\frac{\partial\,}{\partial z}\ket{{\cal E}(z)}=\hat{H}(z)\ket{{\cal E}(z)},
\label{Eq2}
\end{equation}
where
$\ket{{\cal E}(z)}=\sum_{j\in {\cal I}}{\cal E}_j(z)\ket{j}$,
$\ket{j}=(\ldots,0,\stackrel{j\mathrm{-th}}{1},0 ,\ldots )^{\rm t}$, and ${\cal
	I}\subset \mathbb{Z}$ is the set of indices corresponding to  waveguides coupled to each other.
The Hamiltonian $\hat{H}(z)$ reads:
\begin{equation} \label{Hamiltonian}
\hat{H}(z)= f(\hat{n},z)+\left[  g(\hat{n},z)\hat{V}^\dag +
\hat{V}g(\hat{n},z) \right],
\end{equation}
where $\hat{n}\ket{j}=j\ket{j}$ is the number operator and
$\hat{V}\ket{j}=\ket{j-1}$ and
$\hat{V}^\dag\ket{j}=\ket{j+1}$ are step-down and step-up (unitary) phase operators \cite{Louisell,Newton}, not to be confused with Susskind-Glogower operators (non-unitary) phase operators \cite{SG} (see supplementary material).
%
The solution to Eq. (\ref{Eq2}) can be expressed in terms
of the propagator:
\begin{equation}
\ket{{\cal E}(z)}=\hat{U}(z,z_0)\ket{{\cal E}(z_0)},
\end{equation}
which satisfies
\begin{equation}
-{\rm i}\frac{\partial\,}{\partial z} \hat{U}(z,z_0)=\hat{H}(z)\hat{U}(z,z_0)\,,\qquad \hat{U}(z_0,z_0)=I
\label{EqPropagator}
\end{equation}

Note that in the most general case, the Hamiltonian in Eq. (\ref{Hamiltonian}) is $z$ dependent,
thus the propagator $\hat{U}(z,z_0)$ depends on the initial and final values $(z_0,z)$, but in the particular case  of $z$-independent
Hamiltonian, it depends only on the difference $z-z_0$, forming a uniparametric group of operators.

In the general case when some of the system's parameters vary with the propagation distance $z$, closed form solutions are not easy to find. While of course the system can be solved numerically,
analytical or at least semi-analytical solutions can still provide deeper insight. In the following we show how group theoretical techniques can come to aid to achieve this goal.

We can apply Lie's method to find the symmetry algebra of Eq. (\ref{Eq2}). However, since this  is a system of
ordinary differential equations,  they can be characterized by $z$-dependent operators (or matrices for the
case of a finite array) $\hat{I}(z)$, known as
invariant operators, and defined by the condition $\frac{{\rm d}\, }{{\rm d}z}\hat{I}(z)\equiv \frac{\partial\,}{\partial
	z}\hat{I}(z)+{\rm i}[\hat{I}(z),\hat{H}(z)]=0$.
	
If the propagator is known, it is very simple to compute invariant operators. These are given by \cite{Invariants} $\hat{I}(z)=\hat{U}(z) \hat{I}(0)\hat{U}(z)^{-1}$,
where $\hat{I}(0)$ is any constant operator. From this we see that invariant operators have constant eigenvalues and their eigenvectors are
preserved during propagation (i.e. they are solution, up to a suitable phase,
of Eq. (\ref{Eq2})) \cite{{Invariants}}.  They generalize the
\textit{Lewis-Ermakov} invariant operators introduced for the time-dependent quantum harmonic oscillator
\cite{Lewis-Ermakov}. Invariant operators are very useful when studying systems with $z$-dependent
Hamiltonians  (time-dependent in the quantum case), since they provide a \textit{constant framework} to express the dynamics of the system (the Hamiltonian
is no longer valid for this purpose since it is not invariant) \cite{Lewis-Riesenfeld,QAT,QAEPT}.

\subsection{Wei-Norman method}

Consider a discrete photonic systems whose Hamiltonian can be expressed according to the following decomposition (specific photonic arrays that correspond to different scenarios will be discussed shortly):
\begin{equation}
\hat{H}(z)=\sum_{k=1}^N \alpha_k(z) \hat{A}_k,
\end{equation}
where $\alpha_k(z)$ are scalar expansion coefficients and $\hat{A}_k$ are constant matrices (written in terms of the basic operators
$\hat{n},\hat{V}$ and $\hat{V}^\dag$) that close a Lie algebra ${\cal G}$, i.e. $[ \hat{A}_i,\hat{A}_j]=\sum_{k=1}^N c_{ijk}\hat{A}_k$.
Under this condition, the differential equation for the propagator  (\ref{EqPropagator}) can be solved by using the Wei-Norman factorization method \cite{WeiNorman,Vance,Vance2}:
\begin{equation}
U(z,z_0)={\rm e}^{u_1(z,z_0)\hat{A}_1} \ldots {\rm e}^{u_N(z,z_0)\hat{A}_N},
\end{equation}
where the functions $u_k(z,z_0),\,k=1,\ldots N$ satisfy non-linear first-order coupled
differential
equations involving the structure
constants $c_{ij}^{\,\,k}$ and the coefficients $\alpha_k(z)$.
Imposing the initial condition for the propagator
$\hat{U}(z_0,z_0)=I$ leads to the initial conditions $u_k(z_0,z_0)=0\,,k=1,\ldots,N$.

To simplify the notation, we shall assume that $z_0=0$ and it will be omitted in
the sequel, but it should be taken into account that the resulting propagator
$\hat{U}(z)$ can only be applied to the state $\ket{{\cal E}(0)}$ (i.e. there
is no  invariance under translations on $z$).

\subsection{Example: $E(2)$ invariant photonic systems}

Consider the simplest case $f(\hat{n},z)=\alpha_0(z)\hat{n}$, $g(\hat{n},z)=\alpha(z)$, then the Hamiltonian can be written
as
\begin{equation}
\hat{H}(z)= \alpha_0(z) \hat{n}+\alpha(z)( \hat{V} +\hat{V}^\dag)\,.
\end{equation}
These operators close the Euclidean Lie algebra in 2D, ${\cal E}(2)$:
\begin{equation}
[\hat{n},\hat{V}]=-\hat{V}\,,\quad[\hat{n},\hat{V}^\dag]=\hat{V}^\dag\,,\quad[\hat{V},\hat{V}^\dag]=0\,.
\end{equation}
Note that in this case
the lattice is infinite, ${\cal I}=\mathbb{Z}$.

Photonic systems with Euclidean ${\cal E}(2)$ symmetry have been extensively studied, since they correspond to waveguide arrays with uniform coupling (although the coupling can depend on $z$) and a linear propagation constant ramp. These arrays feature Bloch oscillations \cite{Bessel}.

According to the Wei-Norman method, and choosing the order  $+,0,-$ (any order is allowed, but this one leads to a triangular set of coupled differential equations, see Ref. \cite{Wei-Norman-order}), we have
\begin{equation}
\ket{{\cal E}(z)}=\hat{U}(z)\ket{{\cal E}(0)}
={\rm e}^{ u_+(z)\hat{V}^\dag}  {\rm e}^{ u_0(z)\hat{n}}{\rm e}^{ u_-(z)\hat{V}}  \ket{{\cal E}(0)}.
\label{E2}
\end{equation}

Suppose that the Hamiltonian $\hat{H}(z)$ is Hermitian, i.e. $\alpha_0(z)$ and $\alpha(z)$ are
real. This means that the
propagator $\hat{U}(z)$ is unitary, implying conservation of total light
intensity. Imposing $\hat{U}(z)\hat{U}(z)^\dag=I$, the following restrictions
hold:
\begin{equation}
|u_+|=|u_-|\,,\qquad
{\rm e}^{u_0}|u_+|^2=-u_+ u_- .
\label{unitarityE2}
\end{equation}
This implies that there are only three independent real functions to be determined (the real and imaginary parts
of $u_+$, and the relative phase $\lambda(z)$ between $u_-$ and $u_+$).

In this case Eq. (\ref{EqPropagator}) leads to:
\begin{eqnarray}
-{\rm i}\; {u_+}'(z)&=&\alpha(z)+\alpha_0(z) \; u_+(z) ,  \nonumber\\
-{\rm i}\; {u_0}'(z)&=&\alpha_0(z)  , \\
-{\rm i}\; {u_-}'(z)&=& {\rm e}^{u_0(z)}\alpha(z)\nonumber .
\end{eqnarray}
The solutions to these equations are easily found to be:
\begin{eqnarray}
u_0(z)&=& {\rm i} \Phi(z), \nonumber\\
u_-(z) &=&  {\rm i}  \int_0^z {\rm e}^{{\rm i} \Phi(t)}\alpha(t){\rm d}t, \\
u_+(z) &=& -  {\rm e}^{{\rm i} \Phi(z)} u_-^*(z), \nonumber
\end{eqnarray}
with $\Phi(z)=\int_0^z \alpha_0(t){\rm d}t$. These solutions clearly satisfy the restrictions (\ref{unitarityE2}).

For the constant parameters case, i.e. $\alpha(z)=\frac{g}{2}$  and $\alpha_0(z)=\alpha_0$ , the solutions are easily obtained to be:
\begin{equation}
u_0(z)= {\rm i} \alpha_{0} z    \,,\quad  u_+(z)=u_-(z)=\frac{g}{\alpha_{0}}({\rm e}^{{\rm i} \alpha_{0} z}-1) .
\end{equation}
Substituting these solutions into Eq. (\ref{E2}), we obtain the well-known solutions:
\begin{eqnarray}
 \prodesc{m}{{\cal E}_n(z)}&\equiv&\bra{m}\hat{U}(z)\ket{n} \\
&=& {\rm i}^{m-n} {\rm e}^{{\rm i} (m+n) \frac{\alpha_{0} z}{2}}  J_{m-n}\left(\frac{2g}{\alpha_{0}}|\sin(\frac{\alpha_{0}}{2}z)|\right). \nonumber
\end{eqnarray}

\begin{figure}
\center 	\includegraphics[width=0.75\textwidth]{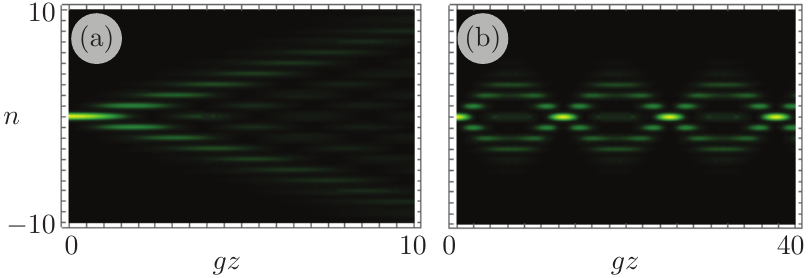}
	\caption{(Color online) Intensity propagation, $|\bra{n}\hat{U}(z)\ket{m}|^2$, through Euclidian waveguide arrays with parameter values (a) $\alpha_{0}(z) = 0 $, $\alpha(z) = g/2$ and (b) $\alpha_{0}(z) = \alpha(z) = g/2$ for light initially impinging at the $m=0$ waveguide.}
	\label{fig:E2}
\end{figure}

In Fig. \ref{fig:E2}a and b we show the intensity propagation through waveguide arrays with Euclidean symmetry for the constant parameters case with $\alpha_{0}=0$  (simulating discrete diffraction in uniform lattices), and $\alpha_{0}=g$, featuring Bloch oscillations \cite{Bessel}.

\subsection{Example: $SU(2)$ invariant photonic systems}

The simplest, finite size photonic system with periodic revivals was introduced in Ref. \cite{Gordon}. The couplings were
designed in order to have equally spaced eigenmodes, resulting in  an underlying ${\cal SU}(2)$
symmetry. Let us discuss the more general case of ${\cal SU}(2)$ symmetry with $z$-dependent
parameters.

In this case $f(\hat{n},z)=\alpha_0(z)(\hat{n}-j)$ and
$g(\hat{n},z)=\alpha(z)\sqrt{\hat{n}(2j+1-\hat{n})}$, where $j$ is a positive
half-integer labeling the irreducible and unitary representations of ${\cal SU}(2)$, with dimension $2j+1$.
Therefore the Hamiltonian can be written as (once transformed to have zero trace):
\begin{equation}
\hat{H}(z)= \alpha_0(z) \hat{J}_0+\alpha(z)( \hat{J}_+ +\hat{J}_-),
\end{equation}
where 
$\hat{J}_0=\hat{n}-j$,
$\hat{J}_+=\sqrt{\hat{n}(2j+1-\hat{n})}\hat{V}^\dag$,
$\hat{J}_-=\hat{V}\sqrt{\hat{n}(2j+1-\hat{n})}$,
closing the Lie algebra of the ${\cal SU}(2)$ Lie group:
\begin{equation}
[\hat{J}_+,\hat{J}_-]=2\hat{J}_0\,,\quad [\hat{J}_0,\hat{J}_\pm]=\pm \hat{J}_\pm .
\end{equation}
Defining $\hat{J}_x=\frac{1}{2}(\hat{J}_++\hat{J}_-)$,
$\hat{J}_y=-\frac{\rm i}{2}(\hat{J}_+-\hat{J}_-)$ and $\hat{J}_z=\hat{J}_0$, the
new matrices generate the ${\cal SO}(3)$ algebra:
\begin{equation}
[\hat{J}_x,\hat{J}_y] = {\rm i}\hat{J}_z \,,\quad
[\hat{J}_y,\hat{J}_z] = {\rm i} \hat{J}_x \,,\quad
[\hat{J}_z,\hat{J}_x] = {\rm i}\hat{J}_y.
\end{equation}
According to the previous section, and choosing again the order  $+,0,-$, we have
\begin{equation}
\ket{{\cal E}(z)}=\hat{U}(z)\ket{{\cal E}(0)}
={\rm e}^{ u_+(z)\hat{J}_+}  {\rm e}^{ u_0(z)\hat{J}_0}{\rm e}^{ u_-(z)\hat{J}_-}  \ket{{\cal
		E}(0)}.
\label{SU2}
\end{equation}

Suppose that the Hamiltonian $\hat{H}(z)$ is Hermitian, i.e. $\alpha_0(z)$ and $\alpha(z)$ are
real.
This means that the
propagator $\hat{U}(z)$ is unitary, implying conservation of total light
intensity. Imposing $\hat{U}(z)\hat{U}(z)^\dag=I$, the following restrictions
hold:
\begin{equation}
|u_+|=|u_-|\,,\qquad {\rm e}^{u_0}|u_+|^2=-u_+ u_-(1+|u_+|^2).
\label{unitaritySU2}
\end{equation}
In this case Eq. (\ref{EqPropagator}) leads to:
\begin{eqnarray}
-{\rm i}\; {u_+}'(z)&=&\alpha(z)\; (1- u_+(z)^2)+\alpha_0(z) \; u_+(z),   \nonumber\\
-{\rm i}\; {u_0}'(z)&=&\alpha_0(z) -2 \alpha(z) \; u_+(z),  \\
-{\rm i}\; {u_-}'(z)&=& {\rm e}^{u_0(z)}\alpha(z). \nonumber
\end{eqnarray}
The first equation is a complex Riccati equation, and the others can be obtained by quadratures once $u_+$ is known.
Imposing the restriction (\ref{unitaritySU2}), there only remain 3 independent
functions, the real and imaginary parts
of $u_+$, and the relative phase $\lambda(z)$ between $u_-$ and $u_+$.

Although solving the complex Riccati equation is not a trivial task, it should be stressed that this has to be done
just once for all the representations of ${\cal SU}(2)$. For the case of
constant parameters, $\alpha(z)=\frac{g}{2},
\,\alpha_0(z)=\alpha_{0}$, and denoting by $\Omega=\sqrt{g^2+\alpha_{0}^2}$, the solutions are:
\begin{equation}
u_+(z)=\frac{g}{\Omega}\tan(\frac{z \Omega}{2})\frac{{\rm i}-\frac{\alpha_{0}}{\Omega} \tan(\frac{z \Omega}{2})}{1+\frac{\alpha_{0}^2}{\Omega^2}\tan^2(\frac{z \Omega}{2})}\,,\,
u_-(z)=u_+(z),
\end{equation}
and thus $\lambda(z)=0$.

Substituting these solutions into Eq. (\ref{SU2}) results into expressions for $\bra{m}\hat{U}(z)\ket{n}$  containing
Krawtchouk polynomials \cite{OrthogonalPolynomials,WolfKravchuk}. A symmetry-based approach, in the form of Gilmore-Perelomov coherent states, for constant waveguide arrays with an underlying $SU(2)$ symmetry was shown in Ref. \cite{GilmorePerelomovCS}, and examples of $z$-dependent cases for  $j=1$, that is three-waveguide couplers, using the Wei-Norman program has been provided in Ref. \cite{RodriguezLara2014p013802}.

\subsection{Example: ${\mathcal{SO}}(2,1)$ invariant photonic $\mathcal{PT}$ symmetric systems}

The case of ${\cal SO}(2,1)$ invariant photonic systems realized unitarity is similar to the ${\cal SU}(2)$ case, with the
difference that the lattice is infinite (either ${\cal I}=\mathbb{Z}$ or ${\cal I}=\mathbb{N}$, see Ref. \cite{GilmorePerelomovCS}
for the case ${\cal I}=\mathbb{N}$, discussed for the constant parameters case). One of the main differences is that
unitarity  in this case implies
\begin{equation}
|u_+|=|u_-|\,,\qquad {\rm e}^{u_0}|u_+|^2=u_+ u_-(1+|u_+|^2).
\label{unitaritySO21}
\end{equation}

We shall discuss, however, the case where ${\cal SO}(2,1)$ is realized non-unitarily in a finite-dimensional representation. In this case, ${\cal SO}(2,1)$ looks very similar to ${\cal SU}(2)$, and the irreducible representations are labeled by
$j$ with dimension $2j+1$.

A simple way of obtaining these non-unitary representations, see Refs. \cite{OL,Symmetry}, is to replace $J_0$ by ${\rm i} J_0$,
and  $J_y$ by ${\rm i}J_y$ to close a Lie algebra, in such a way that the Hamiltonian is:
\begin{equation}
\hat{H}(z)= {\rm i} \alpha_0(z) \hat{J}_0+\alpha(z)( \hat{J}_+ +\hat{J}_-).
\end{equation}

The net effect of this modification is to allow for imaginary values of the refraction index, ${\rm i}\alpha_0(z)$, implying that
the photonic system possesses gain and losses. If the values of the gain and losses are  balanced, the system is PT-symmetric
(we have also
to impose $\alpha(z)$ and $\alpha_0(z)$ to be real and even) \cite{Bender1}.
The Hamiltonian is no longer Hermitian but it is PT-symmetric, $\hat{H}(z)^{\rm PT}=\hat{H}(z)$.

According to the previous example, and choosing again the order  $+,0,-$, we have
\begin{equation}
\ket{{\cal E}(z)}=\hat{U}(z)\ket{{\cal E}(0)}
={\rm e}^{ u_+(z)\hat{J}_+}  {\rm e}^{ u_0(z)\hat{J}_0}{\rm e}^{ u_-(z)\hat{J}_-}  \ket{{\cal
		E}(0)}.
\label{SO21}
\end{equation}
The propagator $\hat{U}(z)$ is no longer unitary, implying that total light
intensity is not conserved along propagation.

In this case Eq. (\ref{EqPropagator}) leads to:
\begin{eqnarray}
-{\rm i} {u_+}'(z)&=&\alpha(z)\; (1- u_+(z)^2)+{\rm i}\alpha_0(z) \; u_+(z),   \nonumber\\
-{\rm i} {u_0}'(z)&=& {\rm i}\alpha_0(z) -2  \alpha(z) \; u_+(z),  \\
-{\rm i} {u_-}'(z)&=& {\rm e}^{u_0(z)}\alpha(z) .\nonumber
\end{eqnarray}
For the case of
constant parameters, $\alpha(z)=\frac{g}{2},
\,\alpha_0(z)={\rm i} \gamma$, and denoting by $\Omega=\sqrt{g^2-\gamma^2}$, the solutions are:
\begin{equation}
u_+(z)=\frac{g}{\Omega}\tan(\frac{z \Omega}{2})\frac{{\rm i}-{\rm i}\frac{\gamma}{\Omega} \tan(\frac{z \Omega}{2})}{1-\frac{\gamma^2}{\Omega^2}\tan^2(\frac{z \Omega}{2})}\,,\,
u_-(z)=u_+(z).
\end{equation}
Substituting these solutions into Eq. (\ref{SO21}) results into expressions
for $\bra{m}\hat{U}(z)\ket{n}$ involving hypergeometric functions \cite{Symmetry}, where three different cases must be considered, namely
$\gamma<g$ ($\Omega\neq 0$ and real), $\gamma=g$ ($\Omega= 0$) and $\gamma>g$ ($\Omega\neq 0$ and pure imaginary).

\begin{figure}
\center \includegraphics[scale=1]{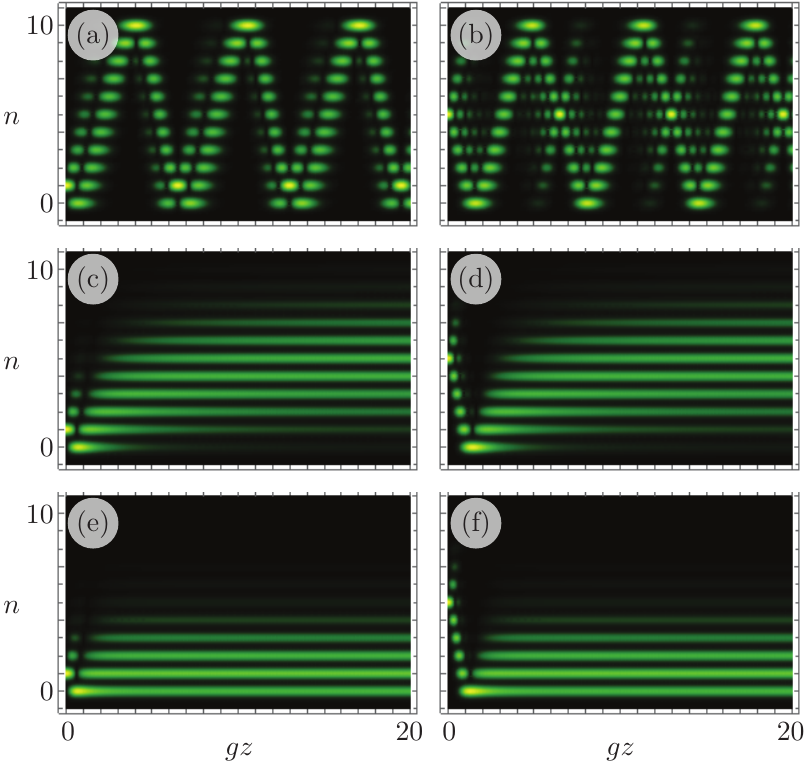}
\caption{(Color online) Renormalized intensity propagation, $|\bra{n}\hat{U}(z)\ket{m}|^2 / \sum_{j}|\bra{j}\hat{U}(z)\ket{m}|^2$, through an $\mathcal{SO}(2,1)$ waveguide array with  $j=5$ and parameter values in
 (a), (b) the $\mathcal{PT}$-symmetric regime,  $\alpha_{0}(z) = \alpha(z)/2  $, and $\alpha(z) = g/2$, (c), (d) Kato's exceptional point,  $\alpha_{0}(z) = \alpha(z) $, and $\alpha(z) = g/2$, and (e), (f) broken $\mathcal{PT}$-symmetric regime, $\alpha_{0}(z) = 3 \alpha(z) / 2 $, and $\alpha(z) = g/2$, for light initially impinging at the $m=1$ (left column) and $m=5$ (right column) waveguides.}
\label{fig:SO21}
\end{figure}

In Fig. \ref{fig:SO21}, plots of $|\bra{n}\hat{U}(z)\ket{m}|^2$ are shown for the case $j=5$, with periodic revivals
 (although non-unitary, i.e. with power oscillations) for $\gamma<g$ (PT-symmetry phase), a power-law growth for $\gamma=g$
(exceptional point), and an exponential growth for $\gamma>g$ (broken PT-symmetry).
For a detailed discussion of the application of group theory to the PT-symmetric case for constant parameters, see Refs. \cite{OL,Symmetry}.

\section{Concluding remarks}

In this article, we presented a brief overview on some of the application of group theory in studying continuous and discrete photonic systems. We did not attempt to review group theory itself, we refer the interested readers to the book by A. Zee \cite{Zee}), but rather to discussed several examples that illustrate the power of this technique. We hope that these carefully selected examples serve as motivation for new research direction that make use of group theoretical methods in advancing new concepts in emerging research areas such as non-Hermitian optics, topological photonics and supersymmetry in optics as well as the connections between some of these different concept \cite{Schomerus1,Topo_non_Hermitian,SUSY_non_Hermitian,SUSY_PT}.

What is important to note here is that group theory can provide an insight that leads to developing general frameworks for dealing with certain problems. Importantly, group theory allows us to categorize and analyze systems possessing the same symmetries in a unified fashion regardless of their dimensionality. For instance, Optical arrays with $SU(2)$ symmetry will exhibit coherent oscillations dynamics that can be represented on the Poincare sphere. Consequently, the behavior of these arrays can be engineered by three independent rotations, an insight that would have been impossible to arrive at without invoking group theory.
Another example is the utilization of group theory to identify a very peculiar family of optical arrays, knowm as Glauber-Fock photonic lattices \cite{GlauberFock}, possesing the same symmetry group as the  harmonic oscillator \cite{GilmorePerelomovCS} that can demonstrate periodic oscillations or a free space-like propagation dynamics, similarly to the $E(2)$ invariant photonic systems discussed previously but
with a semiinfinite array. This interesting feature was discovered by noting that in the symmetry algebra of the harmonic oscillator there are compact operators
with discrete spectrum (featuring periodic oscillations) as well as non-compact operators with continuous spectrum (featuring free propagation). These two examples demonstrate the power of group theory, not only for obtaining new solutions, but also for categorizing photonics systems and predicting new dynamics.
In light of above discussion and due to the increasing complexity of new photonic systems enabled by unprecedented simulation power and dense packing technologies, these new insights derived from group theory are inevitable for studying and designing large photonic systems and building complex optical devices having new functionalities.

\section*{Acknowledgments}

B.M.R.-L. acknowledges support from the Photonics and Mathematical Optics Group at Tecnologico de Monterrey and Consorcio en \'Optica Aplicada through CONACYT FORDECYT $\#$290259 project grant.
R.E. acknowledges support from Henes Center for Quantum Phenomena, Michigan Technological University.
J.G. acknowledges support from Spanish MINECO projects FIS2014-57387-C3-3P and DPI2013-47100-C2-1-P.

\section*{Supplementary material}

To simplify the definitions, let us suppose that the Lie group is a matrix Lie group (i.e. it consists of
invertible matrices of a given dimension $n$).
\\
\textbf{Lie group:} A set $G$ with a binary operation $\circ:\,G\times G\rightarrow G$ and a unary operation ${}^{-1}:\,G\rightarrow G$
such that it is a group (i.e. it is associative, $a\circ (b\circ c)=(a\circ b)\circ c$ 
and there is a neutral element $e$) and $\circ$ and ${}^{-1}$ are differentiable. Matrix Lie groups satisfy
this requirements with the operation $\circ$ being ordinary matrix multiplication.
\\
\textbf{Lie algebra of a Lie group $G$:} It is the set of matrices $X$ (not necessarily invertible) such that $exp(X)$ 
is in $G$. Group multiplication $\circ$ translates into the commutator of matrices $[X,Y]=XY-YX$.
\\
\textbf{Casimir of a Lie algebra:} It is a polynomial C in the Lie algebra (with multiplication being the ordinary matrix
multiplication) such that $[C,X]=0$ for all $X$ in the Lie algebra. The most common Casimirs are quadratic 
Casimirs. For instance for $SU(2)$ the Casimir is $C=J_x^2 + J_y^2+J_z^2$.
\\
\textbf{First order symmetry: } Given a differential equation, a first order symmetry is a first-order
differential operator $L$ such that if $\Psi$ is a solution of the differential equation then $L\Psi$ is also
a solution.
\\
\textbf{Invariant operators: } Given a system characterized by a $z$-dependent Hamiltonian $\hat{H}(z)$,
an invariant operator is a $z$-dependent operator $\hat{I}(z)$ such that if $|{\cal E}(z)\rangle$ is a solution
of the equation $-i\frac{d\, }{dz}|{\cal E}(z)\rangle=\hat{H}(z)|{\cal E}(z)\rangle $ then $\hat{I}(z)|{\cal E}(z)\rangle$  is also a solution. They satisfy 
$\frac{d\, }{dz}\hat{I}(z)\equiv \frac{\partial\,}{\partial z}\hat{I}(z)+i[\hat{I}(z),\hat{H}(z)]=0$.
\\
\textbf{Step-down and step-up phase operators:}  They are exponential-of-the-phase operators, $\hat{V} = \widehat{e^{-i \phi}}\,, \hat{V}^\dag = \widehat{e^{i \phi}}$, i.e. $\hat{V} = \widehat{\cos\phi}-i\,\widehat{sin\phi}\,,\hat{V}^\dag = \widehat{\cos\phi}+i\,\widehat{sin\phi}$.
They were first introduced in quantum mechanics by W. H. Louisell (1963)  and further discussed by R.G. Newton (1980). $\hat{V}$ and $\hat{V}^\dag$ are true unitary operators, provided that the spectrum of the number operator $\hat{n}$ is $\mathbb{Z}$ instead of $\mathbb{N}_0$.
They are similar to the well-known Susskind-Glogower non-unitary phase operators, introduced in (1964) from the usual harmonic oscillator ladder operators $\hat{a}\,,\hat{a}^\dag$, as $\hat{U}=\frac{1}{\sqrt{\hat{N}+1}}\hat{a}$ and $\hat{U}^\dag=\hat{a}^\dag\frac{1}{\sqrt{\hat{N}+1}}$, where $\hat{N}$ is the number operator for the harmonic oscillator, whose
spectrum is $\mathbb{N}_0$. The main difference is that $\hat{V}|0\rangle=|-1\rangle$, whereas $\hat{U}|0\rangle=0$.
\\
\textbf{Compact and non-compact operators:} A compact operator $\hat{K}$ is an operator with discrete spectrum. It satisfies
that the uniparametric group $exp(t\hat{K})\,,t\in\mathbb{R}$ is periodic, thus is it isomorphic to $U(1)$. A non-compact operator $\hat{N}$
is an operator with continuous spectrum. It satisfies that the uniparametric group $exp(t\hat{N})\,,t\in\mathbb{R}$ is not periodic, 
and thus it is isomorphic to $\mathbb{R}$.

\end{document}